# Atomic mechanisms of self-diffusion in amorphous silicon

Matthias Posselt[1*], Hartmut A. Bracht[2], Mahdi Ghorbani-Asl[1], and Drazen Radić[2]

[1]Helmholtz-Zentrum Dresden – Rossendorf, Institute of Ion Beam Physics and Materials Research, 01328 Dresden, Germany
[2]University of Münster, Institute of Materials Physics, 48149 Münster, Germany

**ABSTRACT**

Based on recent calculations of the self-diffusion (SD) coefficient in amorphous silicon (a-Si) by classical Molecular Dynamics simulation [M. Posselt, H. Bracht, and D. Radić, J. Appl. Phys. **131**, 035102 (2022)] detailed investigations on atomic mechanisms are performed. For this purpose two Stillinger-Weber-type potentials are employed, one strongly overestimates the SD coefficient, while the other leads to values much closer to the experimental data. By taking into account the individual squared displacements (or diffusion lengths) of atoms the diffusional and vibrational contributions to the total mean squared displacement can be determined separately. It is shown that the diffusional part is not directly correlated with the concentration of coordination defects. The time-dependent distribution of squared displacements of atoms indicates that in a-Si a well-defined elemental diffusion length does not exist, in contrast to SD in the crystalline Si. The analysis of atoms with large squared displacements reveals that the mechanisms of SD in a-Si are characterized by complex rearrangements of bonds or exchange of neighbors. These are mono- and bi-directional exchanges of neighbors and neighbor replacements. Exchanges or replacements may concern up to three neighbors and may occur in relatively short periods of some ps. Bi- or mono-directional exchange or replacement of one neighbor atom happen more frequently than processes including more neighbors. A comparison of results for the two interatomic potentials shows that an increased three-body parameter only slows down the migration, but does not change the migration mechanisms fundamentally.

* Corresponding author,
Address: Helmholtz-Zentrum Dresden - Rossendorf, Bautzner Landstraße 400,
01328 Dresden, Germany
Electronic address: m.posselt@hzdr.de
Phone: +49 351 260 3279
Fax: +49 351 260 3285

## I. INTRODUCTION

The most fundamental process of atomic transport in solids is self-diffusion (SD). SD in crystalline solids is mediated by native point defects such as vacancies and self-interstitials. In amorphous solids the mechanisms of SD are still not fully understood. The subject of the present work is SD in amorphous Si (a-Si). This material plays an important role in nanoelectronic and display technologies, for solar panels, and in many other applications. For a comprehensive overview on the state of the art on experimental and theoretical research on SD in a-Si the reader is referred to our recent paper [1]. Direct measurements of SD in a-Si have been performed only recently [2,3]. This was due to the difficulty to suppress two processes competing with SD: (i) solid-phase epitaxial recrystallization (SPER) since the a-Si layer is commonly prepared on respective c-Si substrates by deposition or self-ion implantation and (ii) random nucleation of the crystalline phase within a-Si. A further reason was the requirement for very precise analytical techniques based on the use of isotopically modified multilayers and the corresponding depth profiling techniques. In Ref. [1] it was also discussed that only the experimental SD data of Ref. 3 can be considered as reliable since in the other investigation [2] the a-Si sample contained a high level of carbon contamination.

Atomistic computer simulations can contribute to a better understanding of SD. In Ref. [1], the relationship between SD in amorphous silicon (a-Si) and SPER was investigated by classical Molecular Dynamics (MD) simulations. Using various Stillinger-Weber-(SW)-type and Tersoff(T)-type interatomic potentials, it was found that SD as well as SPER can be described by a simple Arrhenius relation and that the activation enthalpies of both processes are rather equal, which is in qualitative agreement with experiments. These results confirmed earlier assumptions that SD and SPER are governed by very similar mechanisms characterized by local bond rearrangements in a-Si. However, for the known SW- and T-type potentials a quantitative agreement with SPER and SD measurements could not be found. Significant improvements could be achieved, if SW-type potentials with an increased value of the three-body parameter were used [1]. This is illustrated in Fig. 1 that shows the SD coefficient determined by MD simulations using the SW potential of Balamane *et al.* [4] as well as a modified version of this potential with a 1.4 times increased three-body parameter. In the following these potentials are called SW10 and SW14, respectively. In Fig. 1 the small circles depict the simulation results [1] and the black line shows the fit to the experimental data of Ref. [3]. Thick lines indicate the temperature range where experiments or simulations were conducted. Due to limitations of computing time, MD simulations were only possible for self-diffusion coefficients above about $10^{-13}$ cm$^2$/s. Therefore, results for SW14 are shown for temperatures higher than the melting point of real existing a-Si. However, it is assumed that the line obtained by the fit to calculated data is also valid at lower temperatures and that this can be used for comparison with measurements. It should be mentioned that even in the case of the SW14 potential a realistic melting behavior can be modeled if a dependence of the three-body parameter on temperature is introduced [1]. However, such a modification is not used here because of the above-mentioned problem regarding computing time at lower temperatures.

The necessity to employ MD simulations in atomic-level studies of SD is due to the fact that in a-Si a lot of different individual activation barriers exist. This is in contrast to single-crystalline materials where the relatively rigid lattice structure limits their number so that diffusion coefficients can be often calculated by more fundamental and more effective methods, where e.g. the energy barrier and the diffusivity are determined by Density Functional Theory and Kinetic Monte Carlo simulations, respectively. Due to the computational effort required for obtaining results, which are comparable with measured data, classical MD simulations must be used in simulations of SD in a-Si. MD based on tight-binding and first-principle schemes would require unrealistically high computing time.

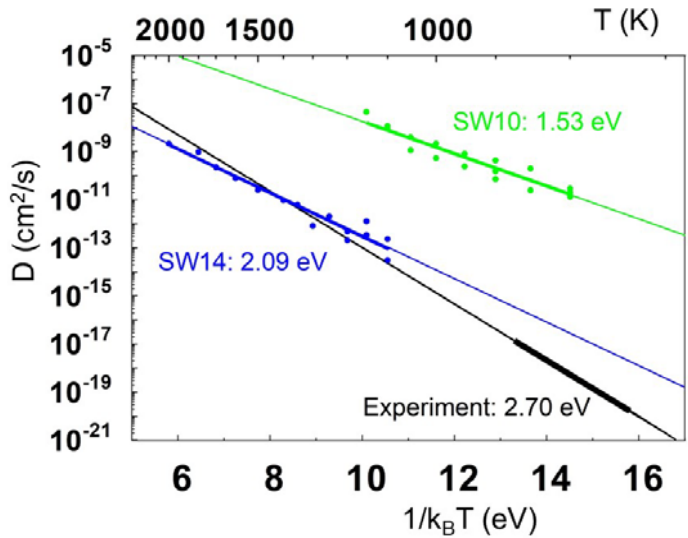

**FIG. 1.** Self-diffusion (SD) coefficient of a-Si determined by recent Molecular Dynamics (MD) simulations [1] using the Stillinger-Weber-type interatomic potential of Balamane *et al.* [4] (SW10) and a modified version of this potential with a 1.4 times increased three-body parameter (SW14). The green and blue lines are the Arrhenius fit to the calculated data depicted by small circles. The black line shows the fit to the experimental data [3]. The temperature range where simulations or measurements were performed are marked by thick lines. The values of the activation enthalpy obtained by the fits are also given.

The present work should be considered as continuation of Ref. [1] and is focused on the atomic-level mechanisms of SD. Results of MD simulations using the SW10 and SW14 potentials are analyzed in the following manner: (i) classification of atoms according to their individual diffusion lengths (or squared displacements) and their corresponding contributions to the SD coefficient, and (ii) identification of most relevant mechanisms of migration, especially for atoms with large diffusion lengths. The two SW-type potentials are considered in order to find out whether the large difference between the calculated SD coefficients is due to different migration mechanisms or is only caused by the changed three-body parameter. Amorphous Si was prepared by MD using (i) melting of diamond-structure c-Si, (ii) equilibration of the melt for several 100 ps, (iii) cooling of the system to 300 K at a cooling rate of 0.1 K/ps, and (iv) equilibration at 300 K for some 100 ps. Also higher cooling rates were investigated and it was found that below 10 K/ps the characteristics of a-Si do not alter significantly. In the preparation of a-Si as well as in SD simulation a cubic cell with 1000 atoms and a MD time step of 1 fs were used. Three-dimensional periodic boundary conditions and the isobaric-isothermal ensemble ($N, P, T$ with $P = 0$) were considered. The simulation cell was coupled to a Berendsen thermostat and a Berendsen barostat [5]. For more details the reader is referred to Ref. [1].

## II. CONTRIBUTION OF ATOMS WITH DIFFERENT DIFFUSION LENGTHS TO SELF-DIFFUSION

The atomic mobility in a-Si is characterized by the time-dependent sum of individual atomic squared displacements $SD_i(t)$ calculated by MD simulations

$$MSD(t) = \frac{1}{N}\sum_{i=1}^{N} SD_i(t) = \frac{1}{N}\sum_{i=1}^{N}[\mathbf{r}_i(t) - \mathbf{r}_i(0)]^2 \qquad (1)$$

where $\mathbf{r}_i(0)$ is the position of atom $i$ at the beginning of the diffusion simulation and $\mathbf{r}_i(t)$ is the position at time $t$, while $N$ is the total number of atoms. In Ref. [1] the self-diffusion coefficient $D$ was obtained from $MSD(t)$ according to the Einstein relation

$$D = \lim_{t \to \infty} \frac{MSD(t)}{6t} \qquad (2)$$

In the present work the atoms were classified during MD simulation regarding to their $SD_i(t)$ over a period of 50 ns. Contributions of atoms with squared displacements greater than or equal to a given threshold, called partial $MSD(t)$, are shown in Figs. 2 (a-d), together with the total $MSD(t)$. It should be noticed that the threshold is given in units of the square of the bond cutoff 2.95 Å. This value corresponds to a position in the region of the broad minimum between the first- and second-neighbor peaks in the pair correlation function (see e.g. [1]). In the following, results for SW10 and SW14 potentials are always compared at a higher temperature (1000 and 2000 K, respectively) and a lower temperature (850 and 1500 K, respectively). At these temperatures the values of total $MSD(t)$ (or the SD coefficient) for SW10 and SW14 are nearly equal (see Fig. 1). It is clear that due to the equal chosen simulation time of 50 ns, all the data for the lower temperature show more statistical fluctuations than those for the higher. In MD simulations, the contribution of atoms with $SD_i(t)$ above or equal a given threshold was determined every 2.5 ns (symbols in the figures), whereas the total $MSD(t)$ was calculated at every MD time step. Figs. 2 (a-d) demonstrate that for each pair of (higher or lower) temperatures the contribution of atoms with squared displacements above a certain threshold is similar, suggesting comparable SD mechanisms in the case of SW10 and SW14. Figs. 3 (a-d) depict the difference between the total $MSD(t)$ and the partial $MSD(t)$. This allows a separation between diffusional and vibrational contributions to the mean squared displacement. The vibrational part, which does not lead to net diffusion and is nearly constant over time, is found for values of partial $MSD(t)$ with a threshold below about 0.1 to 0.2 at the higher temperature and below about 0.05 to 0.075 at the lower temperature. These data correspond to rather large elongations in the vibrational regime (about 1.15 and 0.55 Å, respectively). It should be noticed that the estimation of the precise threshold value from Fig. 3 is difficult due to fluctuation of the corresponding curves. However, this problem has no principal influence on the statement regarding the separation between vibrational and diffusional contribution.

Moreover, the distribution of the individual atomic squared displacements $SD_i(t)$ after different simulation times was determined. Figs. 4 (a) and (b) present the distribution

$$f(SD_i(t)) = \int_0^{SD_i(t)} g(x)dx \qquad (3)$$

Where $g(x)dx$ denotes the probability for a squared displacement within the interval $x...x+dx$. It is clear that the curves change with time due to the general increase of the squared displacement of all atoms. For 1000 K (SW10) and 2000 K (SW14) the overall behavior of the data is similar for the two potentials, which is again an indication of similar migration mechanisms. Also, the distribution defined in Eq. (3) was determined for perfect c-Si. As expected, the corresponding curves are independent of time since in this case atomic displacements are only due to thermal vibrations. Furthermore, c-Si with one vacancy was considered in order to compare with the SD process caused by the presence of this point defect. Fig. 5 depicts the quantity $f(SD_i(t))$ for 1000 K calculated using the SW10 potential.

The smeared steps, in particular for small values of $SD_i(t)$ are an indication of the discrete jumps of the vacancy (or the corresponding lattice atom). The first step is clearly due to the elemental jump length corresponding to the first neighbor distance in c-Si. In contrast, during SD in a-Si many different elemental jump lengths may occur so that the corresponding quantity $f(SD_i(t))$ does not show any steps (see dotted lines in Fig. 5). It should be mentioned that using the SW10 potential, SD in c-Si due to the presence of a vacancy was studied previously [6] and an activation enthalpy of 0.47 eV was found which is much lower than that for SD in a-Si (1.53 eV, see Fig. 1). This also explains the significant difference in the time dependence of the solid and dotted lines in Fig. 5.

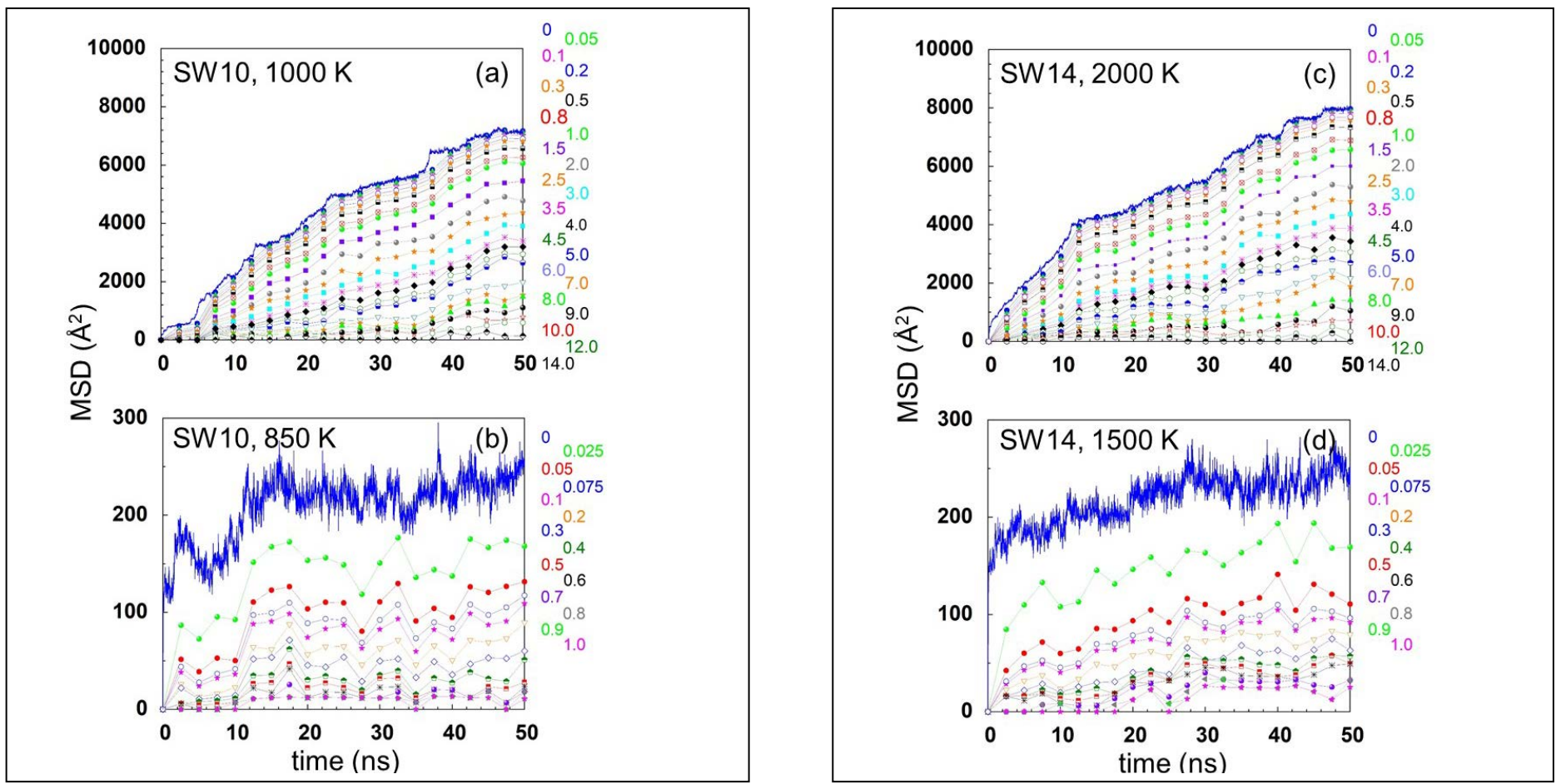

**FIG. 2.** Time dependence of the sum of squared displacements of all atoms $MSD(t)$ (blue curve) as well as the corresponding contributions of atoms with squared displacements above or equal a given threshold (partial $MSD(t)$). The threshold values (right) and the related curves are depicted in different colors. Note that the threshold is given in units of the square of the bond cutoff (2.95 Å). Results of MD simulations using the SW10 and the SW14 potential are shown for 1000 K (a), 850 K (b), 2000 K (c), and 1500 K (d). These temperatures were chosen since at 1000 and 2000 K as well as at 850 and 1500 K the SD coefficients (Fig. 1) determined by the two potentials are nearly equal. For more details see text.

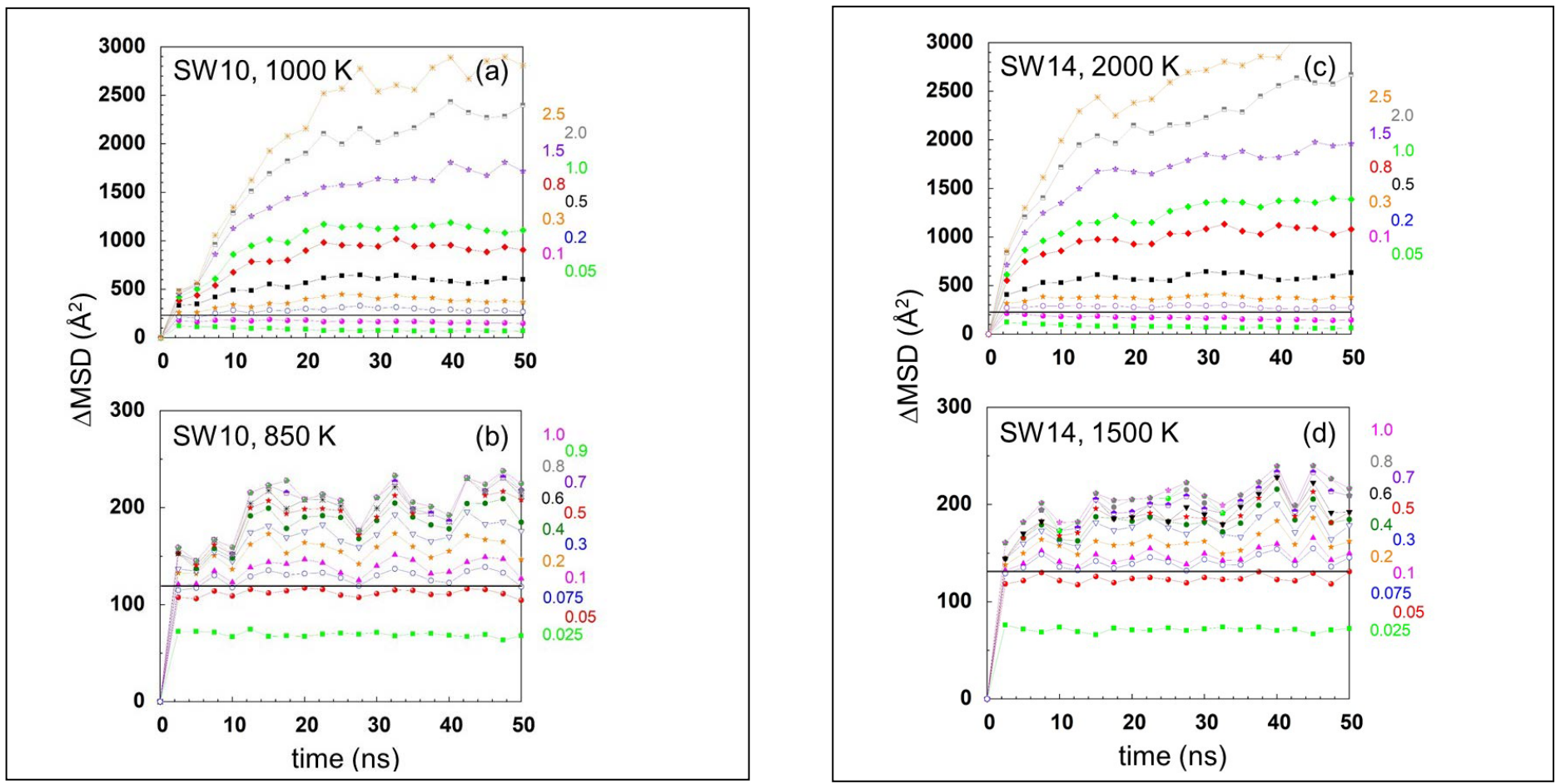

**FIG. 3.** Difference between the total $MSD(t)$ and the corresponding data for given threshold values, i.e. the partial $MSD(t)$ (see Fig. 2). This representation allows a separation between diffusional and vibrational contributions to total $MSD(t)$. The vibrational part, which does not lead to net diffusion, is found for contributions of atoms with a threshold below about 0.1 to 0.2 at the higher temperature, i.e. 1000 and 2000 K, see (a) and (c), and below about 0.05 to 0.075 at the lower temperature, i.e. 850 and 1500 K, see (b) and (d). These values are marked by the horizontal black line.

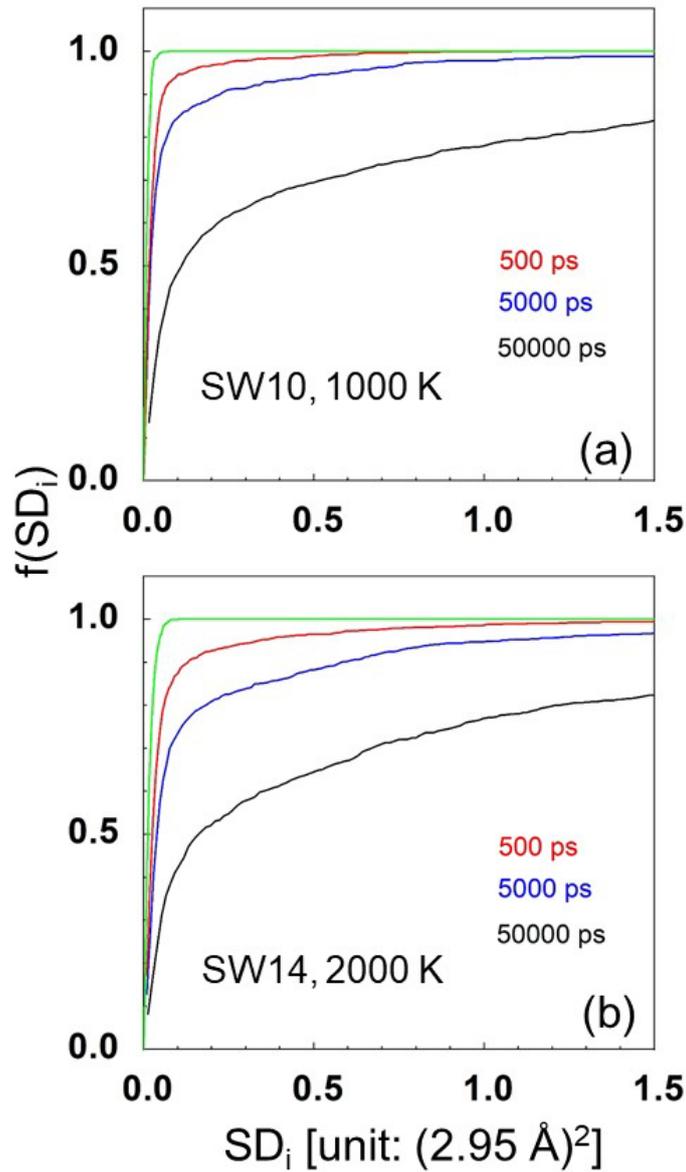

**FIG. 4.** Distributions of individual atomic squared displacements $SD_i(t)$ after a certain simulation time [see Eq. (3)] obtained using SW10 (1000 K) (a) and SW14 (2000 K) (b). Corresponding data determined for perfect single-crystalline Si (c-Si) are depicted by the green line.

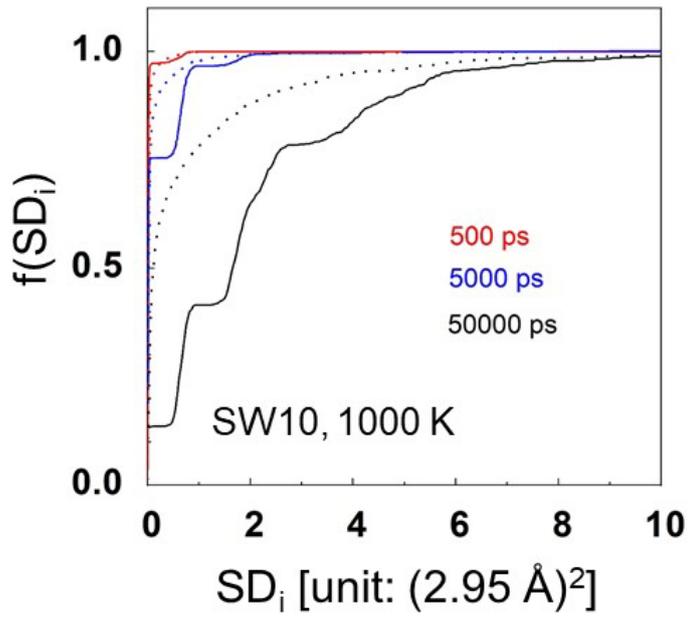

**FIG. 5.** Distributions of $SD_i(t)$ for the case of a perfect c-Si with a single vacancy calculated using the SW10 potential (1000 K). The dotted lines depict the corresponding data for a-Si [see Fig. 4 (a)] for comparison.

## III. ATOMS WITH LARGE DIFFUSION LENGTHS AND THEIR MOST RELEVANT MIGRATION MECHANISMS

Table I shows how many atoms are registered (after 50 ns MD simulation) with an individual squared displacement above a certain threshold. Obviously, in the case of the two SW potentials used in this work the fraction of atoms with squared displacements above 50% of the respective maximum value $SD_i^{\max}$ is only about 1-2% for the higher and 0.3% for the lower temperature. The migration of some of these atoms will be studied in detail below. Comparison of Table I with Fig. 3 allows an estimation of the percentage of atoms contributing to the diffusional part of the mean squared displacement at 50 ns. In the case of the SW10 potential these values are about 45% and 7% at 1000 and 850 K, respectively, and for SW14 55 % and 5% are obtained at 2000 and 1500 K. Since at the respective higher and lower temperature the SD coefficient obtained using SW10 and SW14 has a comparable value (see Fig. 1), this finding again indicates similar migration mechanisms. Comparison with the data on the concentration of three- and fivefold coordinated atoms presented in the supplementary material (Fig. S1) indicates that the above-mentioned percentages are not directly correlated with coordination defects.

In the following the migration of some atoms with large $SD_i$ (bold numbers in Table I characterize their maximum $SD_i$ value) are investigated in detail. Such a choice is reasonable since these atoms are most mobile and should be thus most relevant for SD. As examples, results obtained for SW10 and SW14 at the lower temperatures (850 and 1500 K, respectively) are selected. At the higher temperatures (1000 and 2000 K) the findings are similar to those at lower temperatures but, of course, the migration process is much faster. In general the diffusion mechanisms are characterized by rearrangements of bonds which leads to the exchange of neighbors. Some of them are similar to neighbor exchange processes known from c-Si, such as bond defect formation or dissolution [7-11] and concerted exchange [12]. The subsequent analysis starts with the data for SW14 and 1500 K and is followed by those for SW10 and 850 K.

**Table I.** Result of the analysis concerning the individual squared displacement of atoms (above a certain threshold) after 50 ns MD simulation: (a) for the SW10 potential at 1000 K, (b) for SW10 at 850 K, (c) for SW14 at 2000 K, and (d) for SW14 at 1500 K. The second column shows how many atoms exhibit a squared displacement above the threshold written in column 1. In the third column, individual atom numbers are only given for atoms with large mean squared displacements (time dependence of $SD_i$ is shown in the supplementary material, Figs. S2-S7) since these atoms are considered later in Figs. 6-11. For example, in Table I (d) atoms 94, 535, 541, 179, 320, and 334 are considered in Figs. 6, 7, and 8. Of course, the numbers themselves are only used to illustrate the atomic mechanisms in the particular examples. Cases where atoms reach their maximum $SD_i$ value are marked by bold numbers.

(a)

| $SD_i$ threshold unit: (2.95 Å)² | $N$ | Individual atom numbers (for $N \leq 5$) |
|---|---|---|
| 0 | 1000 | |
| 0.1 | 512 | |
| 0.2 | 413 | |
| 1.0 | 220 | |
| 2.0 | 119 | |
| 3.0 | 76 | |
| 4.0 | 49 | |
| 5.0 | 39 | |
| 6.0 | 25 | |
| 7.0 | 16 | |
| 8.0 | 15 | |
| 9.0 | 11 | |
| 10.0 | 7 | |
| 11.0 | 5 | 195,**437**,777,837,925 |
| 12.0 | 4 | **195**,777,837,925 |
| 13.0 | 3 | 777,837,925 |
| 14.0 | 3 | **777**,837,**925** |
| 15.0 | 1 | **837** |
| 16.0 | 0 | - |

(b)

| $SD_i$ threshold unit: (2.95 Å)² | $N$ | Individual atom numbers (for $N \leq 4$) |
|---|---|---|
| 0 | 1000 | |
| 0.05 | 81 | |
| 0.075 | 54 | |
| 0.1 | 43 | |
| 0.2 | 27 | |
| 0.3 | 13 | |
| 0.4 | 10 | |
| 0.5 | 4 | **419**,599,660,821 |
| 0.6 | 3 | 599,**660**,821 |
| 0.7 | 2 | 599,821 |
| 0.8 | 2 | **599**,821 |
| 0.9 | 1 | 821 |
| 1.0 | 1 | 821 |
| 1.1 | 1 | 821 |
| 1.2 | 1 | **821** |
| 1.3 | 0 | - |

**(c)**

| $SD_i$ threshold unit: (2.95 Å)² | $N$ | Individual atom numbers (for $N \leq 5$) |
|---|---|---|
| 0 | 1000 | |
| 0.1 | 577 | |
| 0.2 | 478 | |
| 1.0 | 230 | |
| 2.0 | 143 | |
| 3.0 | 91 | |
| 4.0 | 60 | |
| 5.0 | 40 | |
| 6.0 | 29 | |
| 7.0 | 21 | |
| 8.0 | 15 | |
| 9.0 | 11 | |
| 10.0 | 8 | |
| 11.0 | 5 | **126**,469,**825**,912,984 |
| 12.0 | 3 | **469**,912,984 |
| 13.0 | 2 | **912,984** |
| 14.0 | 0 | - |

**(d)**

| $SD_i$ threshold unit: (2.95 Å)² | $N$ | Individual atom numbers (for $N \leq 6$) |
|---|---|---|
| 0 | 1000 | |
| 0.05 | 63 | |
| 0.075 | 34 | |
| 0.1 | 28 | |
| 0.2 | 18 | |
| 0.3 | 10 | |
| 0.4 | 8 | |
| 0.5 | 6 | |
| 0.6 | 6 | **94,**179,320,334,**535, 541** |
| 0.7 | 3 | 179,320,334 |
| 0.8 | 3 | **179,**320,334 |
| 0.9 | 2 | 320,334 |
| 1.0 | 2 | 320,334 |
| 1.1 | 2 | 320,334 |
| 1.2 | 2 | 320,334 |
| 1.3 | 2 | 320,334 |
| 1.4 | 2 | **320,334** |
| 1.5 | 0 | - |

For atoms 94, 541, 179, 535, 320, and 334 [see Table I (d**)**] results of MD simulations (over 50 ns) at 1500 K were analyzed regarding to the time dependence of the atomic squared displacement $SD_i$ and the change of neighbors. In the case of atoms 179, 535, 320, and 334 the maximum values of $SD_i$ given in Table I (d) are mainly determined by one important displacement event. In contrast, atoms 94 and 541 perform several small displacements. For details on the time dependence of $SD_i$, the coordination number, and potential and kinetic energy the reader is referred to the supplementary material.

At first (Fig. 6) the bi-directional exchange of one neighbor atom between the nearest neighbors 94 and 541 is considered. Here, one neighbor (atom 955) of atom 94 becomes neighbor of atom 541, and one neighbor (atom 709) of atom 541 becomes neighbor of atom 94, as explained in detail below. This process contributes to the displacements of atoms 94 and 541 registered at the end of the MD simulation, see supplementary material (Fig. S2). Figs. 6 (a-b) illustrate the time dependence of the neighbors of the pair (94,541) within this period. The numbering of atoms which are neighbors of atom 94 or 541 is given in black. Note that there are 1000 atoms in the MD cell. The other numbering (1, 2, 3, 4, 5, see also related colors and symbols) indicates the sequence of distances (to 94 or 541), starting with the smallest. The bond cutoff is always 2.95 Å. Note that the distance of the neighbors to atoms 94 or 541 may slightly change during the simulation so that the neighbor sequence (1, 2, 3, 4, 5) varies over time. Hence, for a given (black) atom number symbols and colors also change. Furthermore, the thermal fluctuations may also lead to a few and very short-time occurrence of other neighbors, which atom numbers are not explicitly given. Figs. 6 (a-b) demonstrate that atoms 94 and 541 are mostly fourfold coordinated within the shown period. At about 49542 ps these atoms exchange one neighbor (709 and 955) bi-directionally. Fig. 6 (c) depicts the scheme of this process that is similar to bond defect formation or dissolution in c-Si [7-11] which is illustrated in Fig. 6 (d) for comparison. Note that in the pictograms or connectivity plots shown in Fig. 6 (c) and in following figures the change of the neighbors of atoms with large squared displacements is shown. In the example of Fig. 6 (c) that means that atoms 94 and 541 *are the moving atoms and their neighborhood is changed due to this migration.* An analysis of the whole time dependence (over 50 ns) of the neighbors of the pair (94,541) shows that most of the displacements of these atoms are caused by a bi-directional exchange of one neighbor (about 230 cases within 50 ns) and a few (7 cases) are due to such an exchange of two neighbors. Most of these neighbor exchanges correspond to "back and forth jumps" where after a while the original state is restored, so that they do not have a net effect on the final value of $SD_i$ (see supplementary material, Fig. S2).

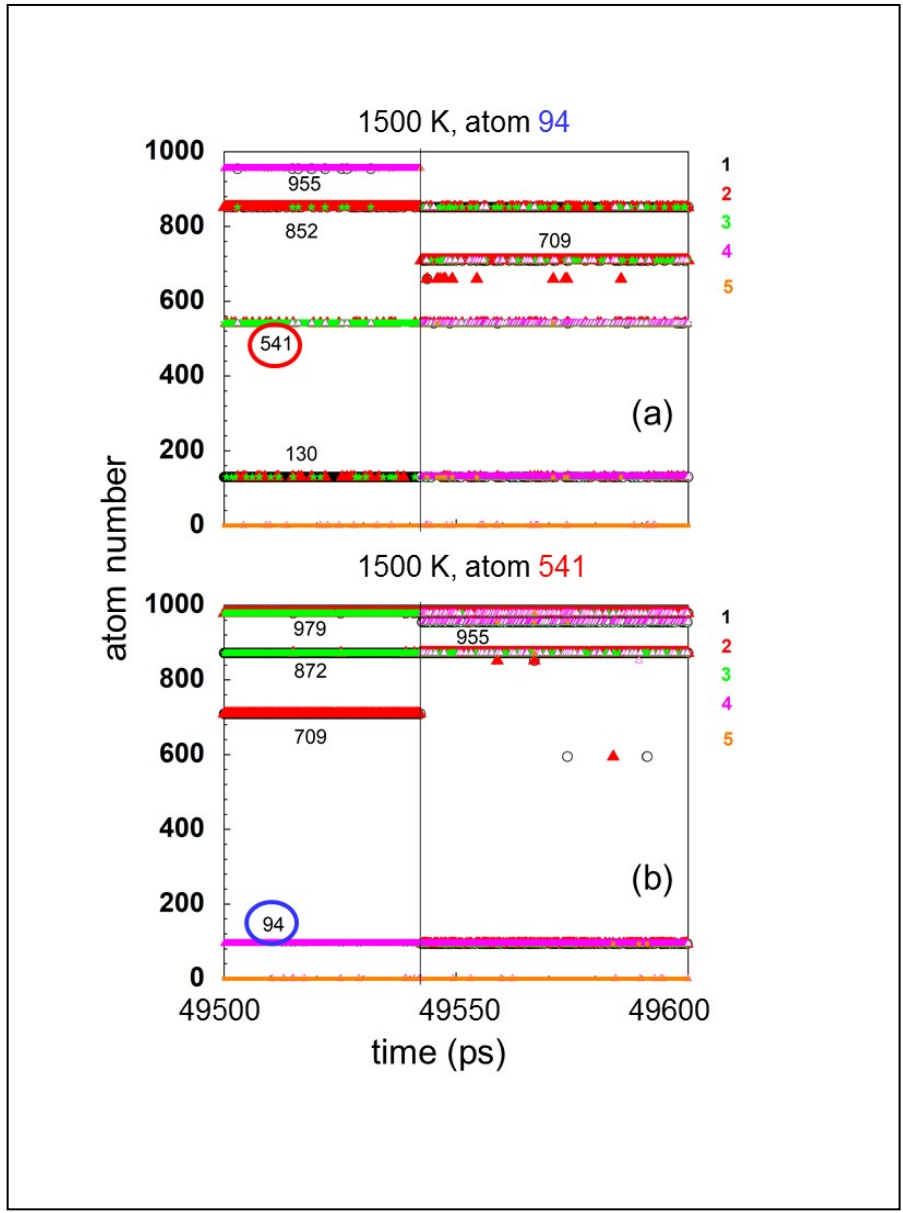
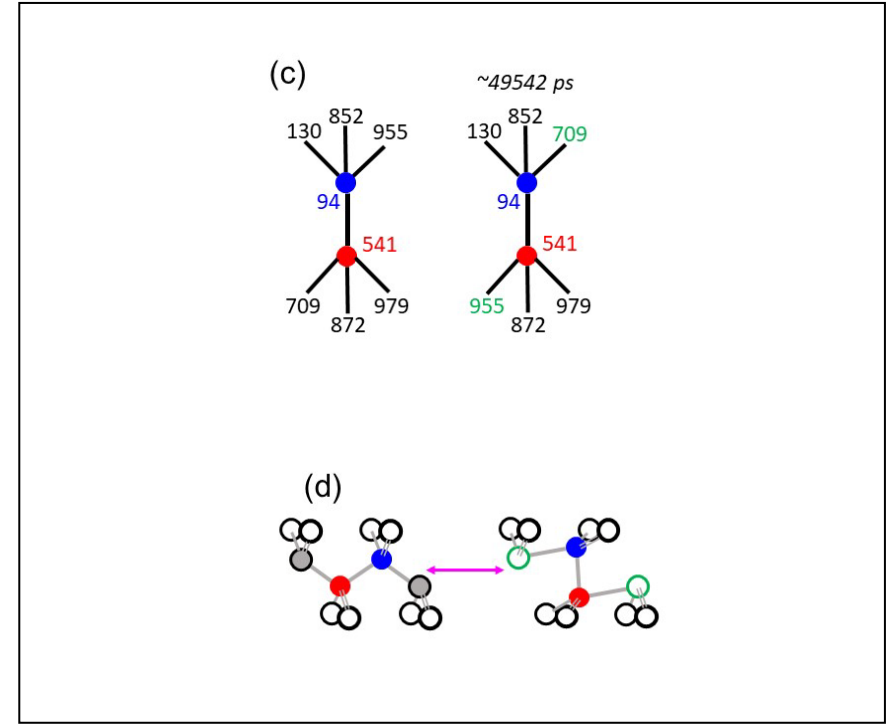

**FIG. 6.** Time dependence of the neighbors of the nearest neighbor atoms 94 (a) and 541 (b). The numbering of neighbors of atom 94 or 541 is given in black. Note that in total there are 1000 atoms in the MD simulation cell. The other numbering (1, 2, 3, 4, 5, see also related colors and symbols in the diagrams) indicates the sequence of distances to atoms 94 or 541, starting with the smallest. The data were obtained by MD simulations at 1500 K using the SW14 potential. The bi-directional exchange of one neighbor between atoms 94 and 541 at about 49542 ps [see (a) and (b)] is summarized schematically in (c). The exchanged neighbor atoms 709 and 955 are marked by green color and bonds are depicted by black lines. The bi-directional neighbor exchange is similar to bond defect formation or dissolution in c-Si, see (d) for comparison.

The second example concerns the bi-directional exchange of all neighbor atoms between the nearest neighbor pair (179,535) observed between 1254.77 and 1257.81 ps. Figs. 7 (a-b) gives an overview on this process which is the only relevant cause for the large permanent $SD_i$ of atoms 179 and 535 within the simulation period of 50 ns (see supplementary material, Fig. S3). The details of the complete neighbor exchange obtained from representations similar to Figs. 7 (a-b), but with higher time resolution, are given in Fig. 7 (c). In c-Si the complete exchange of neighbor atoms between two nearest neighbors is called concerted exchange and was proposed by Pandey [12]. However, due to the rigid lattice structure of c-Si the enthalpy barrier for this process is high so that it does not play any role in practice, since SD in c-Si is mediated by the existing vacancies and self-interstitials. Fig. 7 (c) demonstrates that in a-Si the complete neighbor exchange is possible within a short period. In the relatively flexible amorphous network this process proceeds via a large number of complex intermediate steps where atoms 179 and 535 may be three-, four-, and fivefold coordinated. It must be mentioned that coordination 3 or 5 occurs during the whole simulation time (see data for atoms 179 and 535 given in the supplementary material, Fig. S3), in most cases without any effect. Therefore, there is no unique correlation between the existence of non-fourfold coordinated atoms and their migration. Furthermore, one should keep in mind the role of the farther atomic environment of the atoms depicted in Fig. 7 (c). In these regions smaller but still considerable diffusive displacements of atoms of the flexible a-Si network may occur and may influence the process under investigation. This is different to studies of the migration of a single vacancy or self-interstitial in c-Si where the farther environment of these defects only performs thermal vibrations with a relatively small amplitude.

Furthermore, the relatively large displacements of atoms 179 and 535 at about 29030 ps were studied (see supplementary material, Fig. S3). This is caused by a mono-directional neighbor exchange combined with neighbor replacements (neighbor 27 of atom 179 becomes neighbor of atom 535 while atom 676 occupies the original position of atom 27, and atom 634 leaves the first neighborhood of atom 535), as illustrated in Fig. 7 (d). However, the reverse process occurs at about 29388 ps so that there is no contribution to the final values of atomic squared displacements given in Table I (d).

**FIG. 7.** Bi-directional exchange of three neighbors of the nearest neighbor atoms 179 (a) and 535 (b) within about 3 ps. The data are from simulations at 1500 K using the SW14 potential. For details of the representation (a) and (b) the reader is referred to caption of Fig. 6. With a higher time resolution the schematic representation of the process is given in (c). Exchanged neighbors are shown by green color. Neighbors existing only temporarily are marked by magenta color. For a further explanation of notations and symbols see Fig. 6. The scheme indicates that the bi-directional exchange of all neighbors of the pair (179,535) includes states where atoms 179 and 535 may be three- or fivefold coordinate or are even at a distance greater than the bond cutoff of 2.95 Å. The scheme of the mono-directional neighbor exchange (atom 27) combined with neighbor replacements (27 is replaced by 676, 634 is replaced by 27) at about 29030 ps is shown in (d).

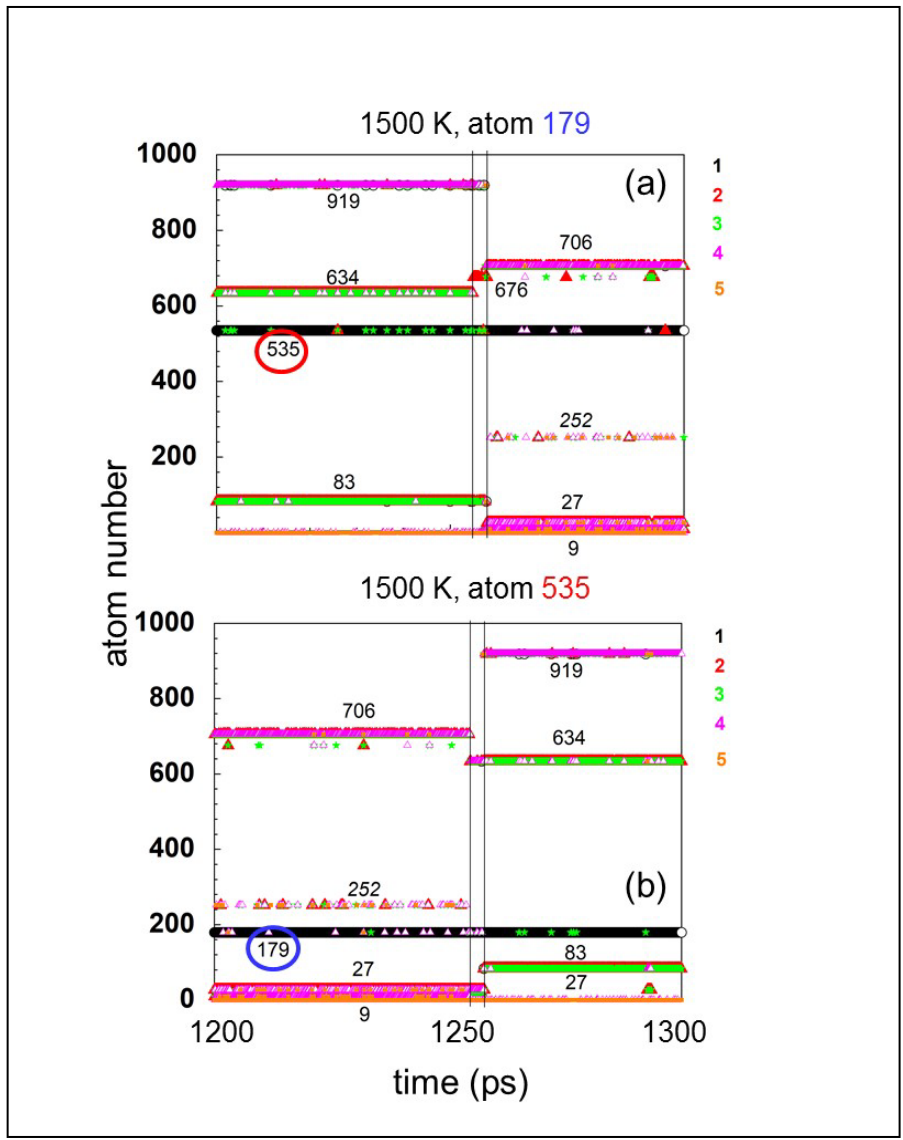
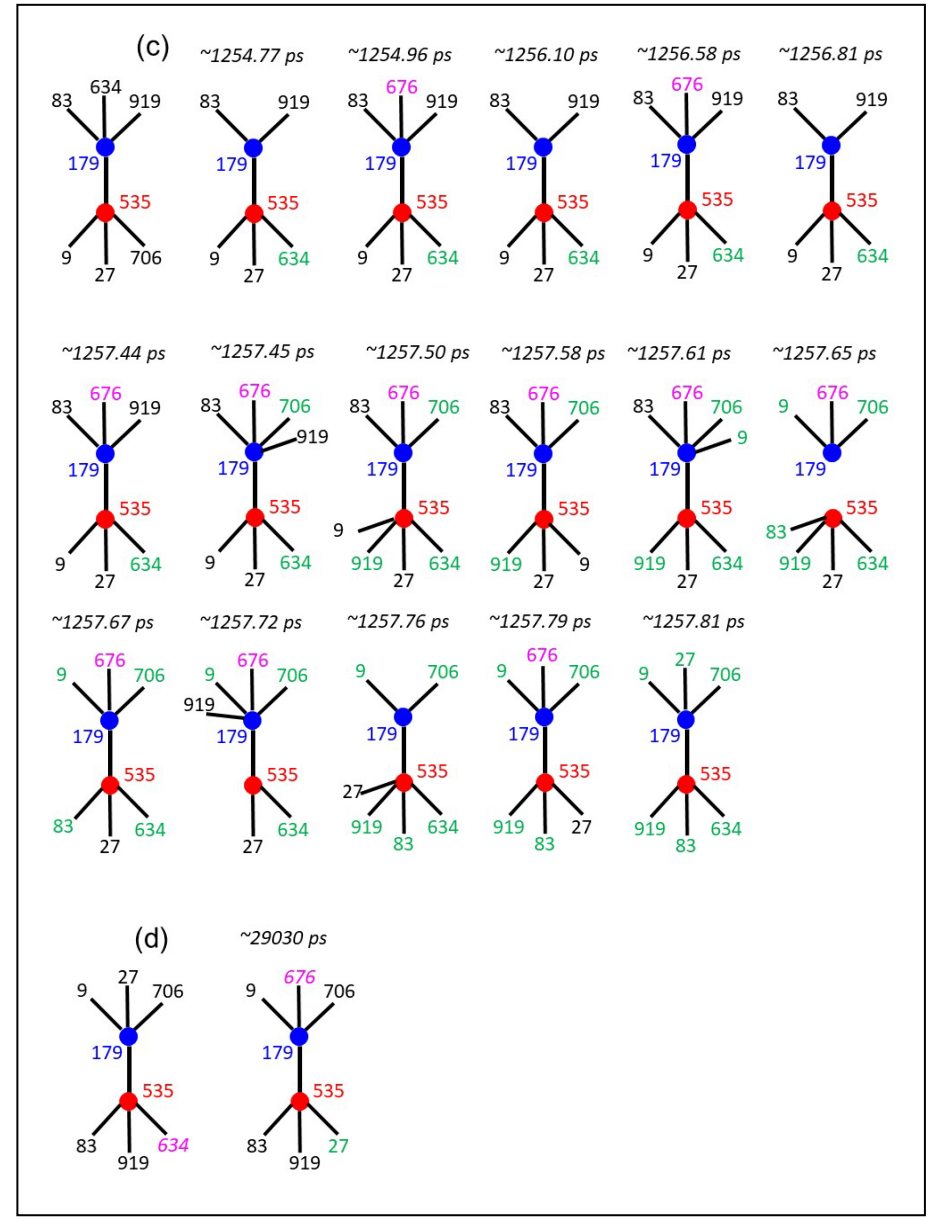

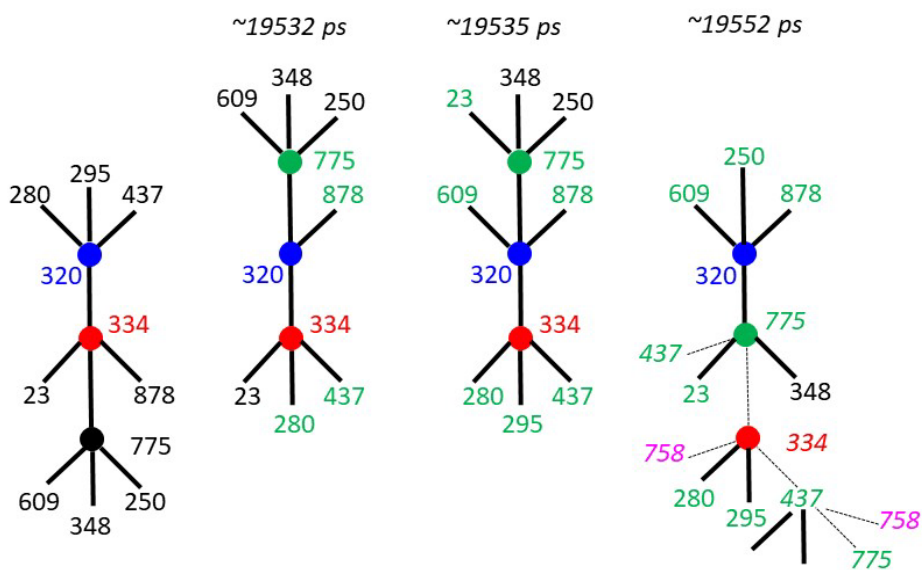

**FIG. 8.** Scheme of the complex rearrangement process during which atoms 320 and 334 experience large displacements within about 20 ps. Also atom 775 is involved, but is not displaced significantly (not shown). The state formed at 19552 ps does more or less persist until the end of the simulations over 50 ns. While atom 320 stays mainly fourfold coordinated, the preferentially fivefold coordinated atoms 334, 775, and also 437, are part of a fluctuating network in which they may appear simultaneously as neighbors of two different atoms. These cases are shown by the thin dashed lines and italic numbers. The depicted data are for the SW14 potential and for 1500 K.

The only large displacements of atoms 320 and 334 occur between about 19532 and 19552 ps (see supplementary material, Fig. S4) and are caused by rather complex neighbor exchanges. Before 19552 ps both atoms are nearest neighbors, in contrast to the time later. The main mechanisms are illustrated in Fig. 8. Within 20 ps the neighbor exchange does not only take place between atoms 320 and 334 but also atom 775 is involved. The complex configuration formed at 19552 ps is more or less characteristic for the remaining time until 50 ns. While atom 320 stays mainly fourfold coordinated, the preferentially fivefold coordinated atoms 334, 775, and also 437, are part of a fluctuating network in which they may appear simultaneously as neighbors of two different atoms. These cases are shown by the thin dashed lines and italic numbers in the last connectivity plot of Fig. 8. It must be noticed that on the time average the $SD_i$ values of all these atoms do not significantly change after 19552 ps (see supplementary material, Fig. S4), but the complex behavior of the amorphous network leads to the above-mentioned fluctuations.

Now the MD results for SW10 at 850 K are analyzed. The time dependence of the atomic squared displacement and the change of neighbors is analyzed for atoms 821, 599, 660, and 419 [see Table I (b)]. The large value of $SD_i$ obtained for atom 821 after 50 ns is caused by two events at about 6927 and 11061 ps (see supplementary material, Fig. S5). The first one consists in a mono-directional neighbor exchange (atom 105) between the nearest neighbor atoms 821 and 374 and the replacement of neighbors (105 by 611, 485 by 729) of atom 821 as shown schematically in Fig. 9 (a). At about 11061 ps, a mono-directional neighbor exchange as well as a replacement/removal of neighbors of atom 821 are observed [see Fig. 9 (b)]. Atom 374 stays fivefold coordinated between 6927 and 11061 ps, and is not significantly displaced throughout the 50 ns (see supplementary material, Fig. S5). As shown in the supplementary material (Fig. S1) , in the temperature region considered, and for SW10, the concentration of fivefold coordinated atoms exceeds that of threefold ones by two order of magnitude, while in the case of SW14 both concentrations are nearly equal and about 1%. This explains the more frequent occurrence of atoms with higher coordination number in figures for the SW10 case.

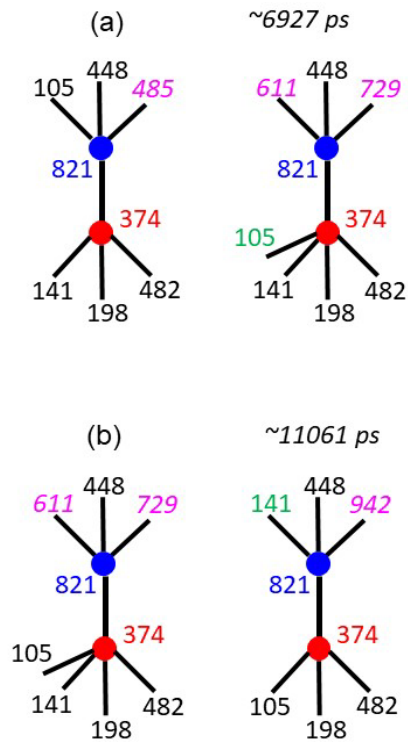

**FIG. 9.** Results from simulations at 850 K using the SW10 potential. For nearest neighbor atoms 374 and 821 both schemes illustrate a mono-directional neighbor exchange and neighbor replacements (italic numbers marked be magenta color). Also the coordination of atom 374 is changed during these processes.

The largest displacement of nearest neighbor atoms 599 and 419 are registered near the end of the MD simulation over 50 ns (see supplementary material, Fig. S6). At about 49128 ps a bi-directional exchange of one neighbor takes place (atoms 194 and 937), which is accompanied by a mono-directional neighbor exchange (atom 6) and an addition of a neighbor (atom 640), see Fig. 10 (a). Before 49128 ps atoms 419 and 599 were four- and fivefold coordinated, respectively. After this time atom 419 is sixfold, and atom 599 fourfold coordinated. At about 49220 ps the bi-directional exchange of two neighbors (marked by green color) occurs, together with a removal of a neighbor (atom 640) of atom 419, so that this atom becomes fivefold coordinated [Fig. 10 (b)]. Details on the time dependence of $SD_i$ of atoms 419 and 599 can be found in the supplementary material (Fig. S6).

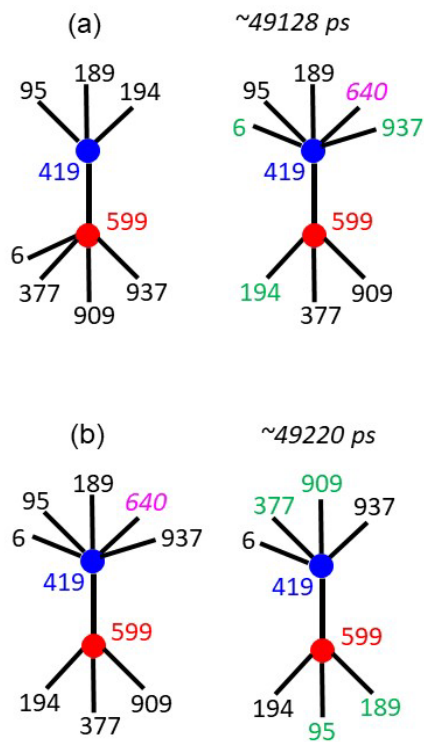

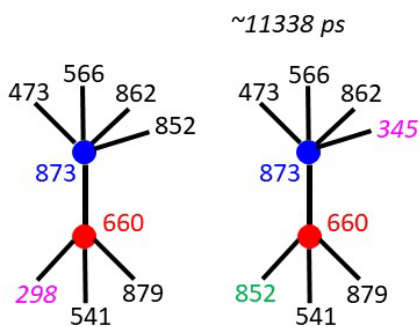

**FIG. 10.** (a) Bi-directional exchange of neighbors 194 and 937 of the nearest neighbor atoms 419 and 599 combined with a mono-directional exchange (6) and a neighbor addition (640). (b) Bi-directional exchange of two atoms (95 and 377, 189 and 909) combined with removal of atom 640. Note the coordination changes during these processes. The results are from simulations at 850 K using the SW10 potential.

**FIG. 11.** Mono-directional neighbor exchange between nearest neighbors 660 and 873 combined with a replacement of neighbors, at about 11338 ps. During this event the coordination numbers of both atoms are not changed. Temperature and interatomic potential used in the simulations are those mentioned in Figs. 9 and 10.

Moreover, the cause for the relative large increase of $SD_i$ of atom 660 at about 11338 ps (see supplementary material, Fig. S7) is investigated. The final (bold) value of $SD_i$ given in Table I (b) is similar to that found after this transition although several "back and forth jumps" are found between 11338 and 50000 ps. The process (Fig. 11) occurring at about 11338 ps consists of a mono-directional neighbor exchange (atom 852) between atoms 660 and 873, and a replacement of a neighbor of atoms 660 (atom 852 is replaced by 345) and 873 (atom 298 is replaced by 852). Finally, it may be worth emphasizing once more that the neighbor changes of the atoms presented in Figs. 6-11 are due their large individual squared displacements as given in Table I and in the supplementary material (Figs. S2-S7).

The results shown in Figs. 6-11 and discussed above are a clear indication that SD in a-Si proceeds via complex rearrangements of atoms or bonds, i.e. bi- or mono-directional neighbor exchange and neighbor replacement. These findings are similar but not equal to those in the very few previous theoretical investigations on these mechanisms [13,14]. Song *et al.* [13] used the nudged elastic band method and the original Stillinger-Weber interatomic potential with a 1.5 times increased three-body parameter to treat so-called perfect events. These are transitions between states in a-Si that directly involve only fourfold coordinated atoms [15]. It should be noticed that this methodology is different from that employed in the present work in which the atomic mechanisms are analyzed "on the fly", i.e. using time-dependent MD data, and no restrictions concerning the coordination number were applied. Song *et al.* [13] found that the most important mechanism is the bond exchange proposed by Wooten *et al.* [7] which corresponds to the bi-directional exchange of one neighbor as discussed above. Also, the bi-directional exchange of two or three neighbors was identified by the authors of Ref. [13]. Interestingly, the mean barrier height obtained for the so-called perfect events is 3.0 eV which is not very different from the experimental value of the activation energy of SD (2.70 eV, [3]). It should be mentioned that the relatively high value of 3.0 eV is due to the high value of the three-body parameter of the potential. An increase of the activation enthalpy of SD in a-Si with increasing three-body parameter was also demonstrated in our recent MD study of SD in a-Si [1]. Furthermore, Santos *et al.* [14] investigated SD in a-Si using Tight-Binding MD simulations and found an unrealistically low activation energy (below 1 eV). In order to identify the atomic mechanism they applied Fast Fourier Transformation (FFT) to filter out thermal vibrations, followed by inverse FFT. Five different processes were identified: (i) bond break, (ii) bond switch, which corresponds to the bi-directional exchange of one neighbor, (iii) frustrated bond switch, which is similar to the mono-directional neighbor exchange combined with a neighbor replacement, (iv) kick-out, and (v) vacancy-like. It is difficult to use the figures depicted in Ref. [14] for a more detailed comparison. Therefore, the processes (i), (iv), and (v) cannot be unambiguously related to a mechanism found in the present work.

## IV. SUMMARY AND CONCLUSIONS

Data obtained from MD simulation of SD in a-Si using the SW10 and the SW14 interatomic potentials were employed to investigate the atomic mechanisms of this process.

The contribution of atoms with different squared displacements (or diffusion lengths) to SD was investigated for a high (1000 and 2000 K) and a low (850 and 1500 K) temperature where the SD coefficients for the SW10 and the SW14 potential are nearly equal. From the results of MD simulations over 50 ns it was found that for each pair of temperatures the contribution of atoms with squared displacements above a certain threshold is comparable, so one may conclude that the SD mechanisms in the case of SW10 and SW14 are obviously very similar. This finding is further confirmed by the estimation of the percentage of atoms contributing to the diffusional part of the mean squared displacement at 50 ns. In the case of the SW10 potential, this value is about 45% and 7% at 1000 and 850 K, respectively, and for SW14 55 % and 5% are obtained at 2000 and 1500 K. These results are not directly correlated to the concentration of coordination defects at these temperatures. The comparison of the time-dependent distribution of the squared displacement of individual atoms with that obtained by SD in c-Si caused by a single vacancy demonstrates that in a-Si a well-defined elemental jump or diffusion length does not exist.

The analysis of atoms with large squared displacements clearly shows that the mechanisms of SD in a-Si are characterized by rearrangement of bonds or exchange of neighbors. Obviously, the increased value of the three-body parameter in the case of SW14 only slows down the migration in comparison to SW10, but does not change the mechanisms fundamentally. These processes can be mainly characterized by mono- and bi-directional exchange of neighbors between nearest neighbor atoms, and by neighbor replacements. The exchanges or replacements may concern one, two, or three neighbors and may occur in relatively short periods of some ps. Bi- or mono-directional exchange or replacement of one neighbor atom are most frequently observed. The bi-directional exchange of one neighbor corresponds to bond defect formation or dissolution in c-Si. If the nearest neighbor atoms are fourfold coordinated, the occurrence of a bi-directional exchange of three neighbors is similar to the concerted exchange in c-Si. While in c-Si (due to existing vacancies and self-interstitials) the mentioned processes do not really contribute to SD, they are highly important for SD in a-Si.

**SUPPLEMENTARY MATERIAL**
The supplementary material contains representations on the temperature dependence of the average concentration of atoms with coordination number 3 and 5, as well as addenda to Figs. 6-11 showing the time dependence of the squared displacements, the coordination number, and the potential and kinetic energy of the corresponding atoms.

**ACKNOWLEDGMENTS**

M. Posselt and M. Ghorbani Asl are grateful for funding by the Deutsche Forschungsgemeinschaft (DFG) via PO436/9-1, H. Bracht and D. Radić for funding by DFG via BR 1520/21-1.

**CONFLICT OF INTEREST**
The authors have no conflicts to disclose.

**DATA AVAILABILITY**
The data that support the findings of this study are available from the corresponding author upon reasonable request.

**SUPPLEMENTARY MATERIAL**

*of the paper*

**Atomic mechanisms of self-diffusion in amorphous silicon**

Matthias Posselt[1*], Hartmut A. Bracht[2], Mahdi Ghorbani-Asl[1], and Drazen Radić[2]

[1]Helmholtz-Zentrum Dresden – Rossendorf, Institute of Ion Beam Physics and Materials Research, 01328 Dresden, Germany
[2]University of Münster, Institute of Materials Physics, 48149 Münster, Germany

* Corresponding author,
Address: Helmholtz-Zentrum Dresden - Rossendorf, Bautzner Landstraße 400,
01328 Dresden, Germany
Electronic address: m.posselt@hzdr.de
Phone: +49 351 260 3279
Fax: +49 351 260 3285

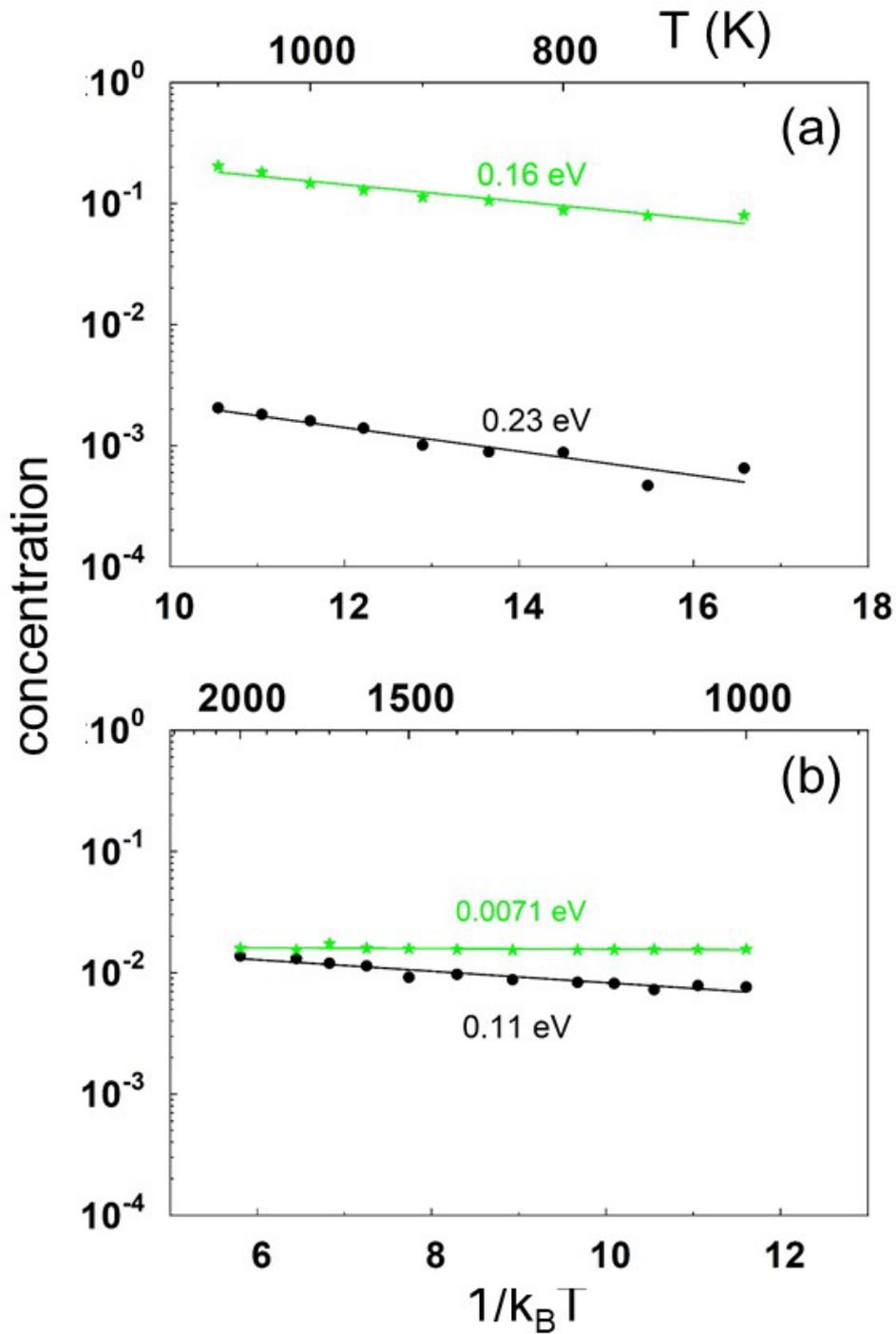

**Fig. S1**
Temperature dependence of the average concentration of atoms with coordination number 3 and 5 obtained by MD simulations using the SW10 (a) and the SW14 potential (b). The activation enthalpies obtained from the Arrhenius fit to the simulation data are also given.

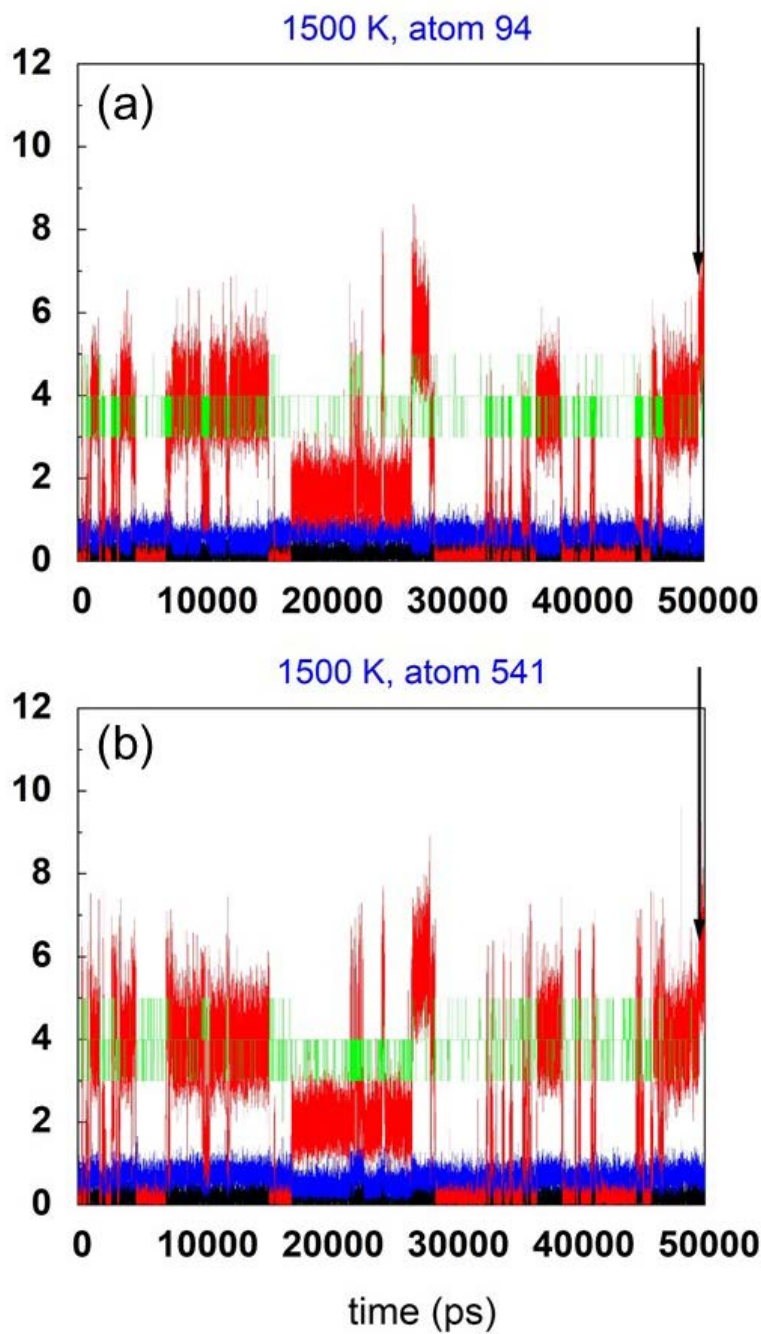

**Fig. S2: Addendum to Fig. 6**

Time dependence of $SD_i$ (red), of the coordination number (green), and of potential (blue) and kinetic (black) energy of atoms 94 (a) and 541 (b), shown over the whole simulation time of 50 ns. The black arrow indicates the transition illustrated in **Fig. 6**. The blue curve is obtained by subtracting the ground state energy in c-Si (-4.63 eV, as a reference value) from the atomic potential energy. The numbers on the ordinate correspond to $SD_i$ values in Å², energy values in eV and the dimensionless coordination number.

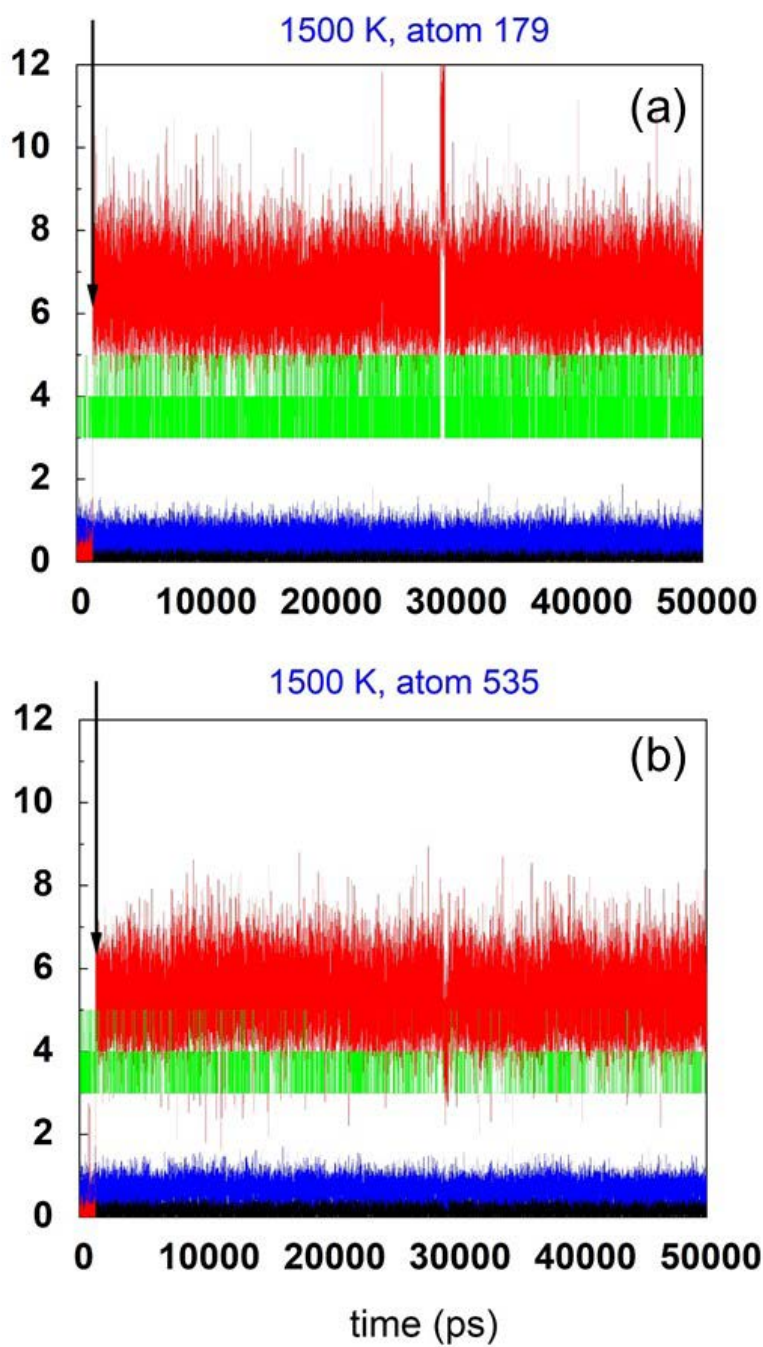

**Fig. S3: Addendum to Fig. 7**

The formal representation is identical to **Fig. S2** but this figure illustrates the case of atoms 179 (a) and 535 (b) and the black arrow marks the transition depicted in **Fig. 7**. This transition which is responsible for the relatively large value of $SD_i$ found after 50 ns (see also **Table Id**). Moreover, the displacement at 29030 ps and the reverse process at 29388 ps are also visible.

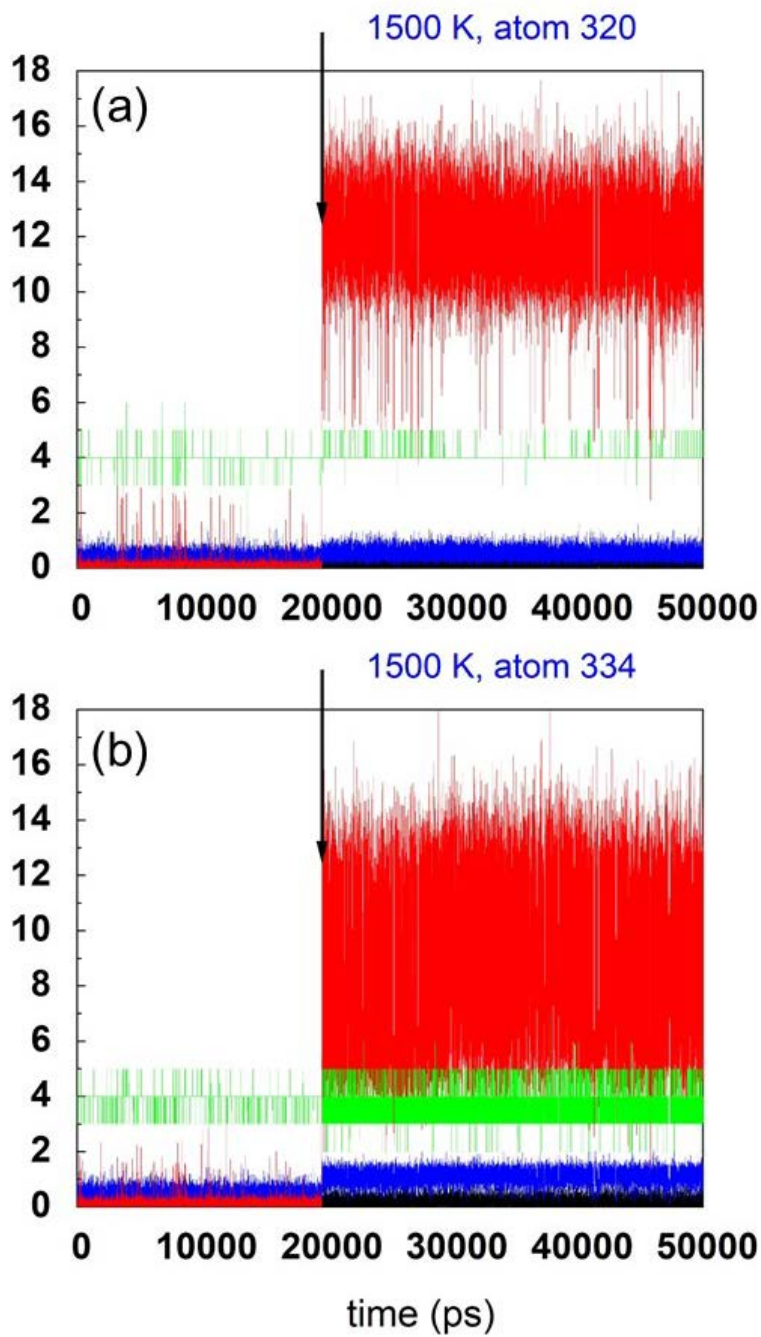

**Fig. S4: Addendum to Fig. 8**
Characteristics (see **Fig. S2**) of atoms 320 (a), 334 (b), and 775 (c) vs. time. The process between 19532 and 19552 ps described in **Fig. 8** is indicated by the arrow.

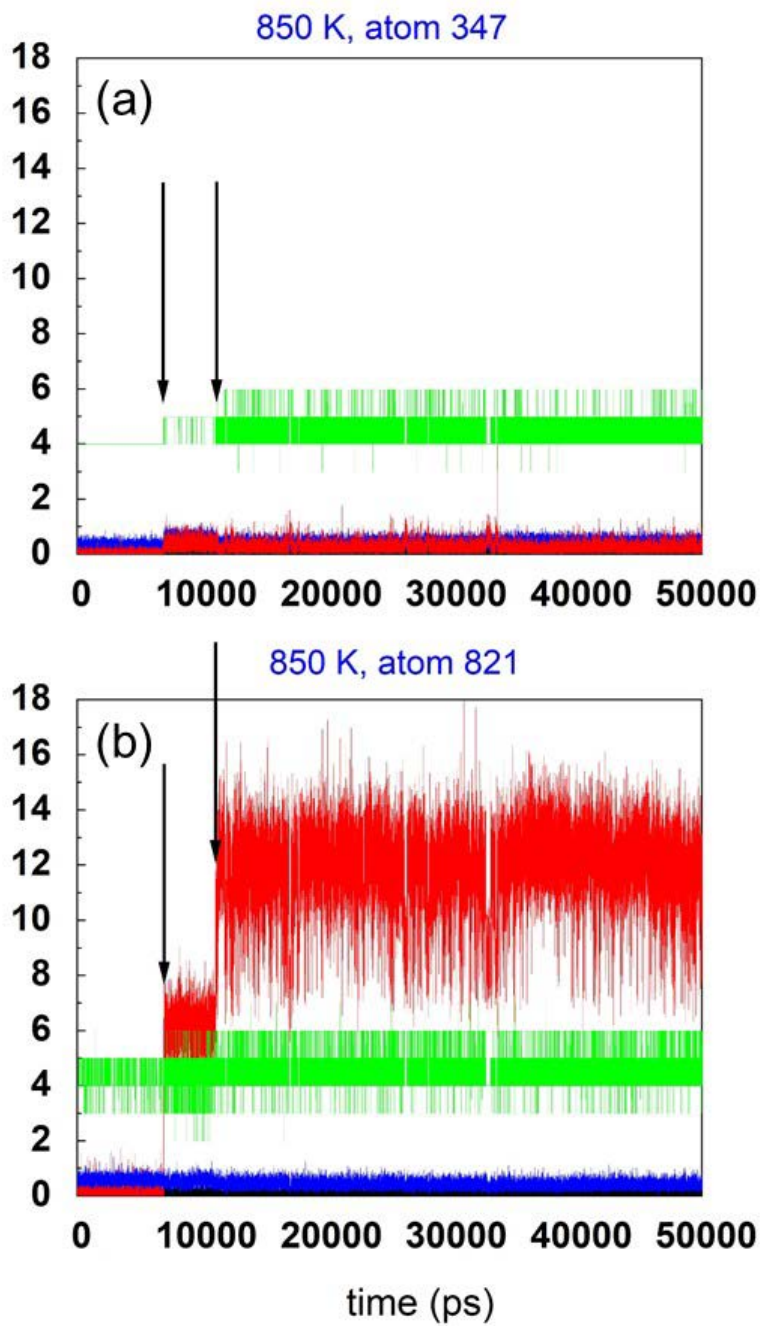

**Fig. S5: Addendum to Fig. 9**
Characteristics of atoms 374 (a) and 821 (b). The time of the transitions illustrated in **Figs. 9 (a-b)** is marked by arrows.

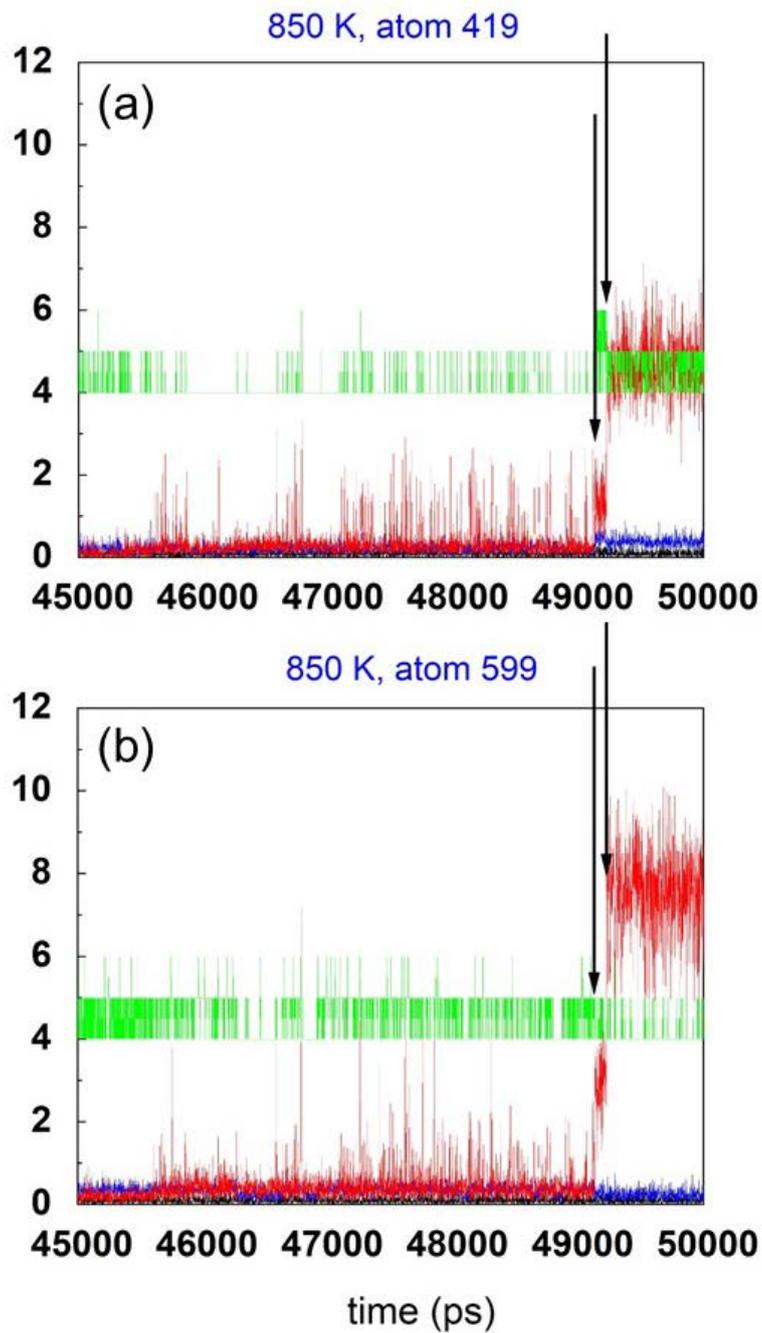

**Fig. S6: Addendum to Fig. 10**
Characteristics of atoms 419 (a) and 599 (b) for the period between 45000 and 50000 ps. The events illustrated in **Figs. 10 (a-b)** are marked by arrows.

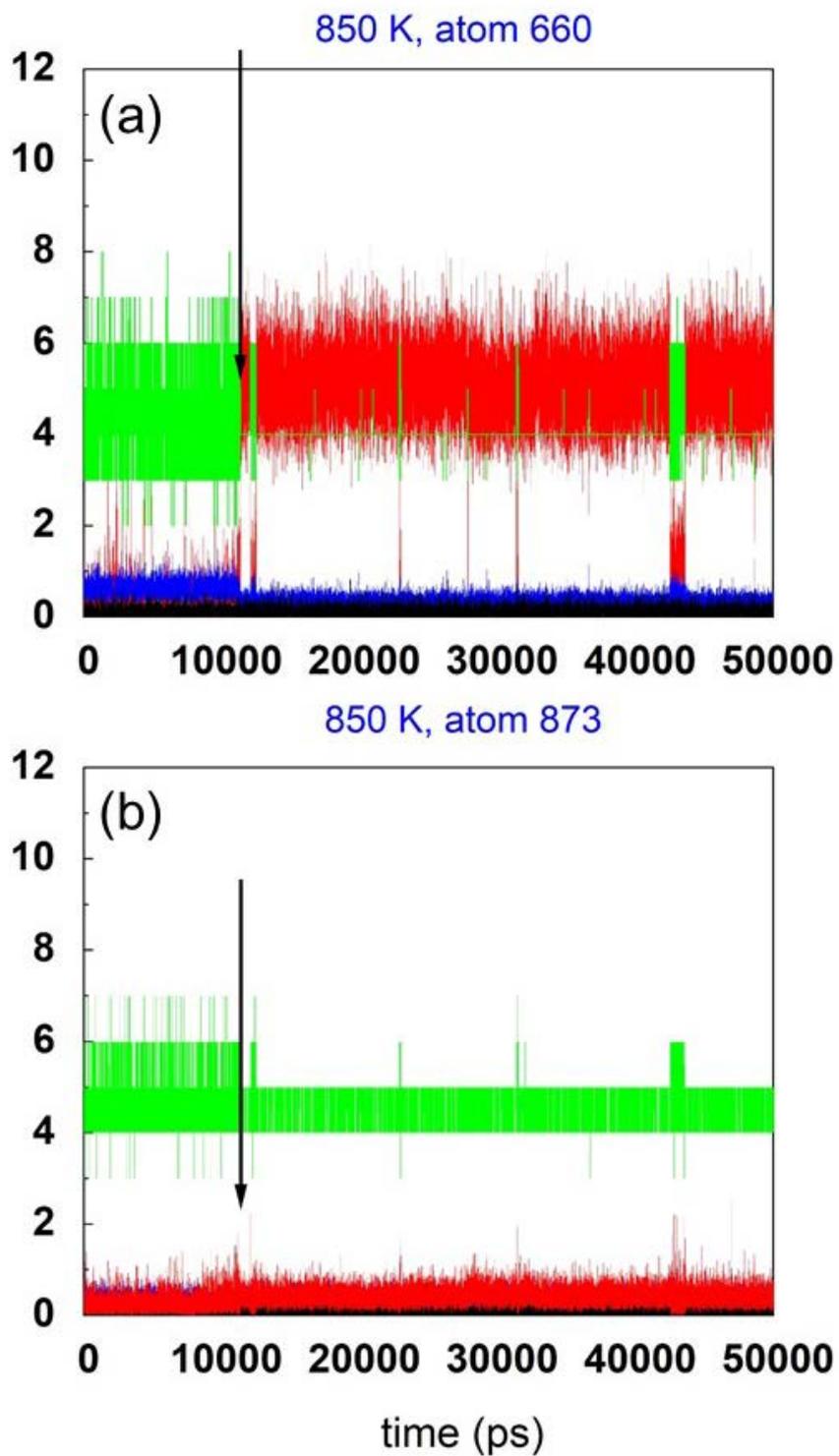

**Fig. S7: Addendum to Fig. 11**
Characteristics of atoms 660 (a) and 873 (b). The arrow indicates the transition shown in **Fig. 11.**